\documentclass[twocolumn,showpacs,amsmath,amssymb]{revtex4}

\usepackage[T1]{fontenc}
\usepackage{graphicx}
\usepackage{dcolumn}
\usepackage{bm}

\input epsf
\renewcommand{\vec}[1]{\mbox{\boldmath $#1$}}

\begin{document}

\preprint{}

\title{
Fission barriers in neutron-proton isospin plane for heavy neutron-rich nuclei
}

\author{F. Minato}
\author{K. Hagino}

\affiliation{
Department of Physics, Tohoku University,
Sendai 980-8578, Japan}

\date{\today}
%
\begin{abstract}

We discuss the sensitivity of fission barrier for heavy neutron-rich 
nuclei to fission paths in the two dimensional neutron-proton
quadrupole plane. To this end, 
we use the constrained Skyrme-Hartree-Fock + BCS method, and 
examine the difference of fission
barriers obtained with three constraining operators, that is, 
the neutron, proton, and mass quadrupole operators. 
We investigate $^{220}$U, $^{236}$U, and $^{266}$U, 
that is relevant to r-process nucleosynthesis. 
We find that the fission barrier heights are almost the same among 
the three constraining operators even for neutron-rich nuclei,
indicating that the usual way to calculate fission barriers with the mass
quadrupole operator is well justified. 
We also discuss the difference between proton and neutron deformation
parameters along the fission paths. 
\end{abstract}
\maketitle
\section{Introduction} 

Fission plays a decisive role in determining the stability of heavy
nuclei, where the Coulomb energy competes with 
the nuclear surface energy. 
Typical examples are superheavy elements (SHE). 
Although the fission barrier disappears in SHE in the liquid drop
model, the nuclear shell effect leads to a relatively 
high fission barrier and eventually stabilizes SHE. 
The experimental efforts have been continued in 
many facilities to synthesize SHE using heavy-ion fusion reactions 
\cite{fus06}.

It has been well recognised that fission plays an important role also 
in the context of nuclear
astrophysics\cite{SFC65,Kodama},
but systematic investigations on its role in r-process 
nucleosynthesis have started only in recent
years \cite{Role.fis.r-pro.,FissionCycling, Imp.fis.}. 
The r-process is one of the most promising candidates that synthesize 
the elements heavier than iron (Fe). 
In this model, nuclei capture 
a number of 
neutrons via 
successive $(n,\gamma)$ reactions in a highly 
neutron-rich environment {\it e.g.,} neutron stars.  
As a consequence, 
the r-process path passes through the neutron-rich side in 
nuclear chart which cannot be reached experimentally at this moment. 
Heavy neutron-rich nuclei produced by the r-process 
may decay by 
spontaneous fission, neutron-induced fission, or beta-delayed fission 
\cite{Role.fis.r-pro.,FissionCycling,Kodama, Imp.fis.}. 
The neutrino-induced fission might also play a role if the neutrino 
flux is significant \cite{KLF04}. 
In order to 
construct a reliable r-process model with fission, 
it is urged to calculate systematically fission barriers 
of many neutron-rich nuclei. 

Theoretically, fission barriers can be calculated using either 
the macroscopic-microscopic model \cite{moller} or 
microscopic mean-field models 
\cite{Sys.fis.bar.,HFB, V.ExtensiontoFission}. 
In the latter approach, one selects a few important degrees of freedom for
fission, such
as quadrupole or higher multipole moments, and draws a fission 
energy surface using the constrained Hartree-Fock method with the
corresponding constraining operators. The total energy is minimized
with respect to all the other degrees of freedom 
than those considered explicitly in the calculation. In this sense, 
the mean-field approach provides an adiabatic potential energy surface 
for the case where the selected degrees of freedom are much slower 
than the other degrees of freedom so that they adiabatically follow
the motion of the former at every instant. 
 
Usually, one takes a mass ({\it i.e.,} proton+neutron) 
quadrupole moment as one of the most 
important degrees of freedom. This implicitly assumes either that the
isoscalar motion is much slower than the isovector motion 
or that the isoscalar and isovector motions are decoupled. 
For fission of neutron-rich nuclei, however, it is not obvious whether
this assumption is justified, and 
it may be more natural that the shape degrees of freedom for 
neutron and proton are treated separately. 
In fact, a two dimensional energy surface spanned by proton and
neutron deformations has been drawn recently for light neutron-rich nuclei,
such as $^{16}$C and $^{20}$O\cite{NWLthesis, QpQnfor16C, 20O}.
The difference in neutron and proton deformation parameters along a
fission path for actinide nuclei 
has also been investigated in Refs. 
\cite{Dobrowolski,GHFB}. 

In this paper, we examine 
the sensitivity of fission barriers 
for neutron-rich nuclei 
to a choice of constraining 
operator in the isospin plane.  
To this end, we use the constrained Skyrme-Hartree-Fock + BCS method, 
and consider the mass, proton, and neutron quadrupole moments as the 
constraining operators. The constrained Hartree-Fock method with the
proton (neutron) constraint provides an adiabatic energy surface for 
the case where the proton (neutron) is much slower than neutron
(proton). 

The paper is organised as follows. 
In Sec. II, we use a schematic model and illustrate an example in which the
constrained-Hartree-Fock method with the three different constraints 
give significantly different results from each other. 
In Sec. III, we briefly summarize the theoretical framework for
constrained Skyrme-Hartree-Fock method. 
Sec. IV presents the results 
for the fission barrier, single particle levels, and 
the proton and neutron deformations along the fission paths 
for uranium isotopes. 
We then summarize the paper in Sec. V. 

\section{Schematic Model}

In the constrained Hartree-Fock method with a constraining operator 
$\hat{O}$, one minimizes the expectation value of 
\begin{equation}
\hat{H}'=\hat{H}-\lambda \hat{O}, 
\label{CHamiltonian}
\end{equation}
where $\hat{H}$ is the Hamiltonian of the system. The Lagrange
multiplier $\lambda$ is determined 
so that the expectation value of $\hat{O}$ 
becomes a given value $O_0$. 
The $\hat{O}$ can be any one-body operator, but usually the mass 
quadrupole operator, $\hat{Q}=\hat{Q}_p+\hat{Q}_n$, 
where $\hat{Q}_p$ and $\hat{Q}_n$
are the quadrupole operators for proton and neutron, respectively, 
is considered as one of the constraining operators $\hat{O}$ 
when one studies fission barriers. 
The aim of this paper is to compare such fission barriers with those obtained 
by using $\hat{O}=\hat{Q}_p$ or $\hat{Q}_n$. 
We call the latter scheme ``proton (or neutron) constraint'' while 
the former ``total constraint''. 

\begin{figure}
\begin{center}
\includegraphics[width=0.45\linewidth,clip]{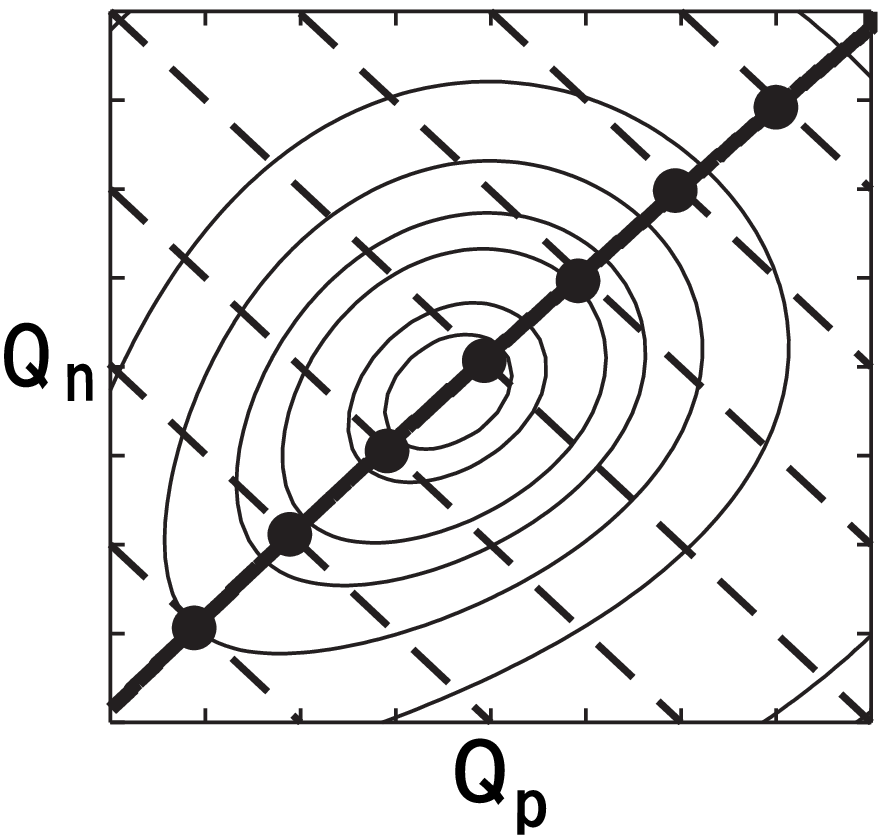}
\includegraphics[width=0.45\linewidth,clip]{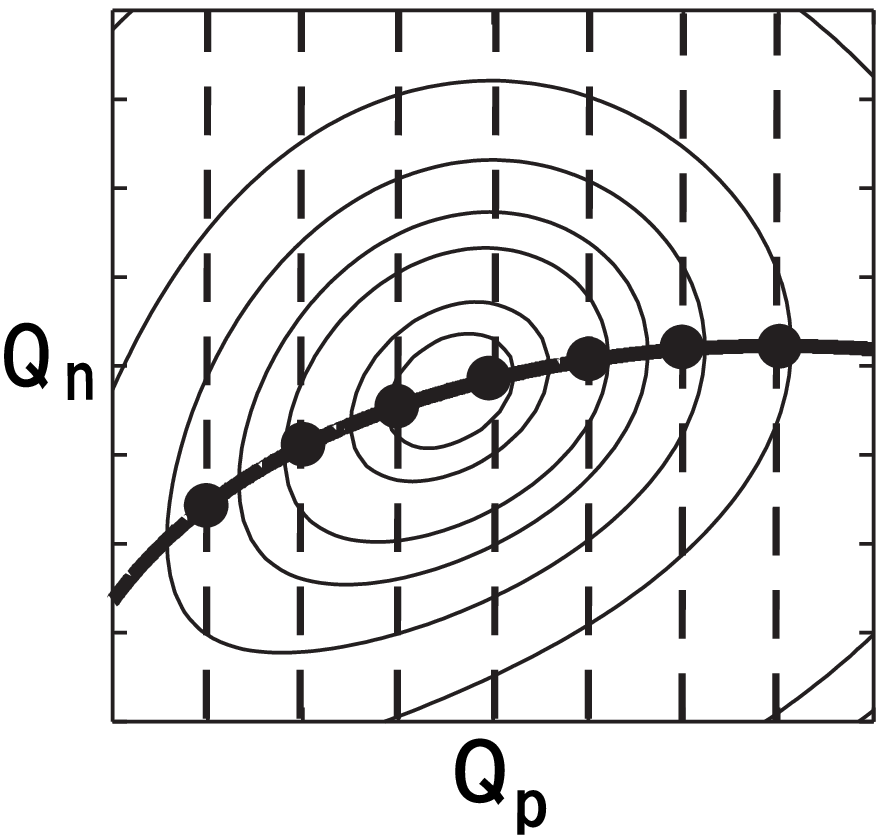}
\caption{
Schematic pictures of the constrained Hartree-Fock method.
The left panel is for the total constraint, in which 
the energy is minimized 
along the lines of $Q_p+Q_n=const.$ shown by the dashed line.
The right panel is for the proton constraint, in which 
the energy is minimized 
along the 
lines of $Q_p=const$. 
The corresponding paths are shown by the thick solid lines in both the 
figures. 
}
\label{fig-schematic}
\end{center}
\end{figure}

Before we perform self-consistent calculations, we would like to 
illustrate a possible difference among the three schemes for the 
constrained Hartree-Fock using a 
schematic model. 
Suppose that we have an energy surface shown in Fig. 1 
in the two-dimensional plane of proton and neutron quadrupole moments, 
$Q_p$ and $Q_n$.
In the total constraint scheme, 
the energy minimum is searched along the line
$\langle Q \rangle =\langle Q_p \rangle + \langle Q_n \rangle = const.$, 
which is shown by the dashed lines in the left panel. 
The resultant path is denoted by
the thick solid line, and the energy variation along this path is shown in
Fig. 2 by the solid line. 
In the case of proton constraint, on the other hand, 
the energy minimum is searched along 
the dashed lines in the right panel of Fig.1,  
which correspond to $\langle Q_p \rangle = const$. 
The path and the energy are shown by the thick solid line in
Fig. 1 and the dotted line in Fig. 2,
respectively. The energy is plotted as a function of the total quadrupole
moment along the path. 
Those of the neutron constraint are obtained in a similar way. 

\begin{figure}
\begin{center}
\includegraphics[width=0.70\linewidth,clip]{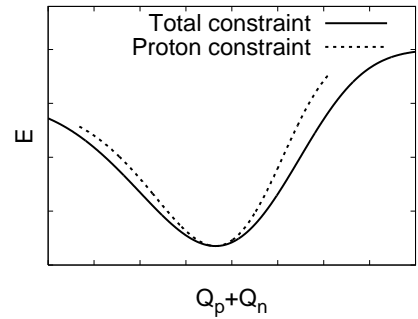}
\caption{
The energy along the paths shown in Fig. \ref{fig-schematic}. 
The solid and the dotted lines are for the total and the proton
constraints, respectively. 
}
\label{fig-schematicbarrier}
\end{center}
\end{figure}

We see clearly that the two paths 
obtained with the different constraining operators 
deviate significantly from each other. 
The energy is also different as a function of the total quadrupole
moment, although the absolute minimum can be obtained irrespective to 
the choice of the schemes. 
The ambiguity arises when the number of degree of freedom is reduced 
from two to one. 

Notice that the differences among the schemes will be 
small if the energy surface is 
much steeper along the line of   
$\langle Q_p \rangle + \langle Q_n \rangle = const$. 
In the next section, we will investigate how much the fission barriers
are changed for realistic nuclei 
depending on which 
scheme one employs to minimize the energy.

\section{Numerical details}

In order to calculate fission barriers for realistic nuclei, we use 
the Skyrme-Hartree-Fock+BCS method \cite{VB72} (see Ref. \cite{BHR03}
for a recent review). 
In this method, the expectation value
of the Hamiltonian $H$ is given in terms of an energy functional as 
\begin{equation}
E=\int d\vec{r} \, \mathcal{H}(\vec{r})
\end{equation}
with 
\begin{equation}
\begin{split}
\mathcal{H}(\vec{r})
&
=\frac{\hbar^2}{2m}\tau(\vec{r})\\
&
+\frac{1}{2}t_0\Big((1+\frac{1}{2}x_0)\rho^2-(x_0+\frac{1}{2})(\rho_n^2+\rho_p^2)\Big)\\
&
+\frac{1}{24}t_3\rho^\alpha\Big((2+x_3)\rho^2-(2x_3+1)(\rho_p^2+\rho_n^2)\Big)\\
&
+\frac{1}{8}\Big(t_1(2+x_1)+t_2(2+x_2)\Big)\tau\rho\\
&
+\frac{1}{8}\Big(t_2(2x_2+1)-t_1(2x_1+1)\Big)(\tau_p\rho_p+\tau_n\rho_n)\\
&
+\frac{1}{32}\Big(3t_1(2+x_1)-t_2(2+x_2)\Big)(\vec{\nabla}\rho)^2\\
&
-\frac{1}{32}\Big(3t_1(2x_1+1)+t_2(2x_2+1)\Big)\Big((\vec{\nabla}\rho_p)^2+(\vec{\nabla}\rho_n)^2\Big)\\
&
-\frac{1}{16}(t_1x_1+t_2x_2)\vec{J}^2+\frac{1}{16}(t_1-t_2)(\vec{J}_n^2+\vec{J}_p^2)\\
&
+\frac{1}{2}W_0(\vec{J}\cdot\vec{\nabla}\rho+\vec{J}_p\cdot\vec{\nabla}\rho_p+\vec{J}_n\cdot
\vec{\nabla}\rho_n)\\
&
+\mathcal{H}_C(\vec{r}). 
\end{split}
\label{Skyrme}
\end{equation}
Here, $\rho_q(\vec{r})$, $\tau_q(\vec{r})$, and $\vec{J}_q(\vec{r})$
are the nucleon density, the kinetic energy density, and the 
spin density, respectively,
which are defined as,
\begin{equation}
\begin{split}
\rho_q(\vec{r})&=\sum_{i\in q,\sigma} v_i^2 \,| \phi_i(\vec{r},\sigma,q) |^2,\\
\tau_q(\vec{r})&=\sum_{i\in q,\sigma} v_i^2 \,| \vec{\nabla}\phi_i(\vec{r},\sigma,q) |^2,\\
\vec{J}_q(\vec{r})&=(-i)\sum_{i\in q,\sigma,\sigma'} v_i^2 \,\phi_i^*(\vec{r},\sigma,q)
\Big( \vec{\nabla}\phi_i(\vec{r},\sigma',q)\times\langle\sigma|\vec{\sigma}|\sigma'\rangle \Big).
\end{split}
\end{equation}
In these equations, $q$ denotes the isospin ($q$=p or n), $\phi_i$ is
the single-particle wave function, and 
$v_i^2$ is the occupation probability estimated in the BCS
approximation. 
$\mathcal{H}_C(\vec{r})$ in Eq. (\ref{Skyrme}) is the Coulomb energy
term, while 
$\rho$, $\tau$, and $\vec{J}$ are the total (proton+neutron) densities.

In this paper, we use the quadrupole operator 
\begin{equation}
\hat{Q}_q =\sqrt{\frac{16\pi}{5}}\sum_{i\in q} r_i^2Y_{20}(\theta_i),
\end{equation}
as a constraining operator in Eq. (\ref{CHamiltonian}). 
For simplicity, 
we assume the reflection and axially symmetric nuclear shapes, 
although the mass asymmetry sometimes plays an important role in
describing nuclear fission. 
From the expectation value of the quadrupole operator, 
we calculate the total deformation parameter as \cite{HLY06}
\begin{equation}
\beta_t=\sqrt{\frac{5}{16\pi}}\frac{4\pi}{3AR_0^2}\langle Q_t \rangle, 
\end{equation}
where $R_0$ is the nuclear radius parameter given by $R_0=1.1A^{1/3}$
(fm).
The proton and neutron deformation parameters are given by
\begin{equation}
\begin{split}
\beta_p&=\sqrt{\frac{5}{16\pi}}\frac{4\pi}{3ZR_0^2}\langle Q_p \rangle\\
\beta_n&=\sqrt{\frac{5}{16\pi}}\frac{4\pi}{3NR_0^2}\langle Q_n \rangle,
\label{defparameter}
\end{split}
\end{equation}
respectively. 

In the actual numerical calculations shown in the next section, we use
the computer code {\tt SKYAX} \cite{skyax,RDN99}. This code solves the 
Skyrme Hartree-Fock equations in the coordinate space with 
the reflection and axially symmetries. 
We use the mesh size of 0.6 fm. 
The pairing correlation is taken into account in the BCS 
approximation. 
In this paper, 
we use the delta force 
\begin{equation}
v_{\rm pair}(\vec{r},\vec{r}')=-V_0\,\delta(\vec{r}-\vec{r}'), 
\end{equation}
for the pairing interaction. 
We expect that our conclusion is qualitatively the same even if 
we use a density-dependent delta interaction. 
In the code, the smooth cut-off function 
\begin{equation}
f_\alpha=\frac{1}
{1+\exp\left(\left(\epsilon_\alpha-\lambda-\Delta E\right)/\mu\right)},
\end{equation}
is introduced for the pairing active
space. Here, $\lambda$ is the Fermi energy, and $\Delta E$
is determined so that 
\begin{equation}
N_{\rm{act}}=\sum_{\alpha}f_\alpha=N_q+1.65N_q^{2/3},
\end{equation}
with $\mu=\Delta E/10$, $N_q$ being the number of particle for proton
($q=p$) or neutron ($q=n$). 
We use the strength parameter of $V_0=$ 279.082 MeV$\cdot$ fm$^3$ for 
proton and 258.962 MeV $\cdot$ fm$^3$ for neutron 
pairings\cite{RDN99}.

\section{RESULTS}

We now present the results of constrained Hartree-Fock calculation 
for the fission barriers of $^{220,236,266}\rm{U}$ nuclei. 
The $^{236}\rm{U}$ is on the $\beta$-stability line, while 
$^{220}\rm{U}$ and $^{266}$U are proton-rich and neutron-rich nuclei,
respectively. Notice that 
$^{266}$U is relevant to r-process nucleosynthesis \cite{r-process-path}.
We adopt the parameter set SLy4 \cite{SLy4} for the Skyrme
functional. 

\begin{figure}
\begin{center}
\includegraphics[width=0.98\linewidth,clip]{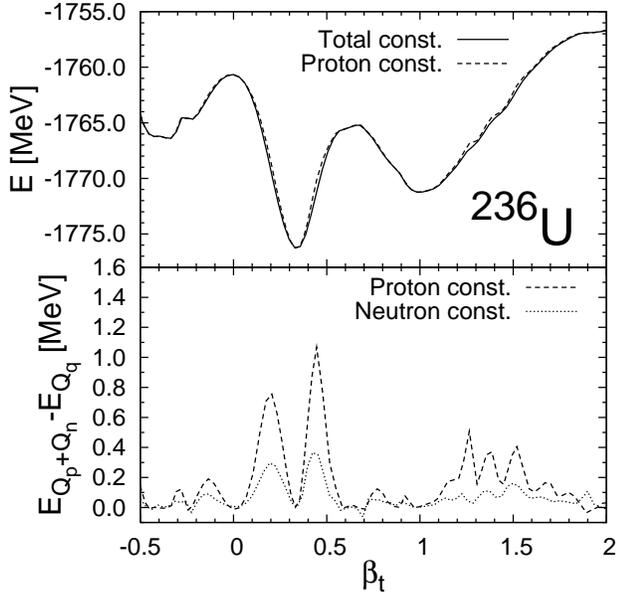}
\caption{
The fission barrier for ${}^{236}\rm{U}$ 
as a function of deformation parameter $\beta_t$ obtained with the 
total constraint scheme (the upper panel).
The dashed line in the 
lower panel shows the difference of the fission barrier obtained with 
the proton constraint and that with the total constraint, while the 
dotted line denotes the difference obtained with the neutron
constraint. 
}
\label{fig-fission}
\end{center}
\end{figure}

\begin{figure}
\begin{center}
\includegraphics[width=0.98\linewidth,clip]{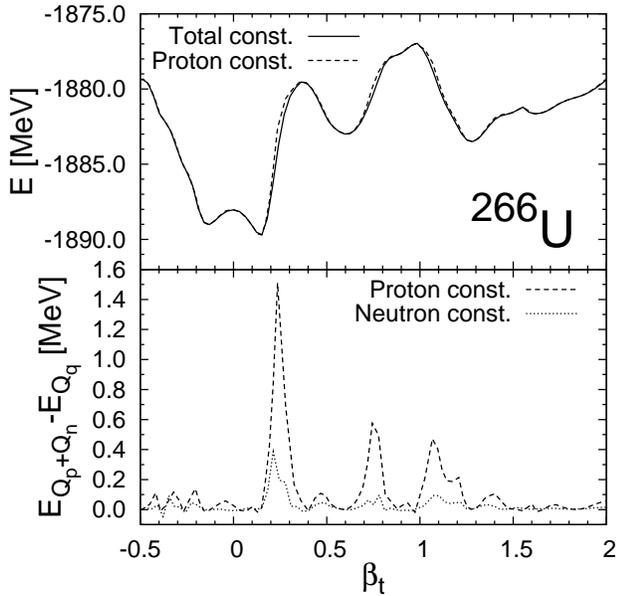}
\caption{
Same as Fig. \ref{fig-fission}, but for 
${}^{266}\rm{U}$. 
}
\label{fig-fission2}
\end{center}
\end{figure}

\begin{figure}
\begin{center}
\includegraphics[width=0.98\linewidth,clip]{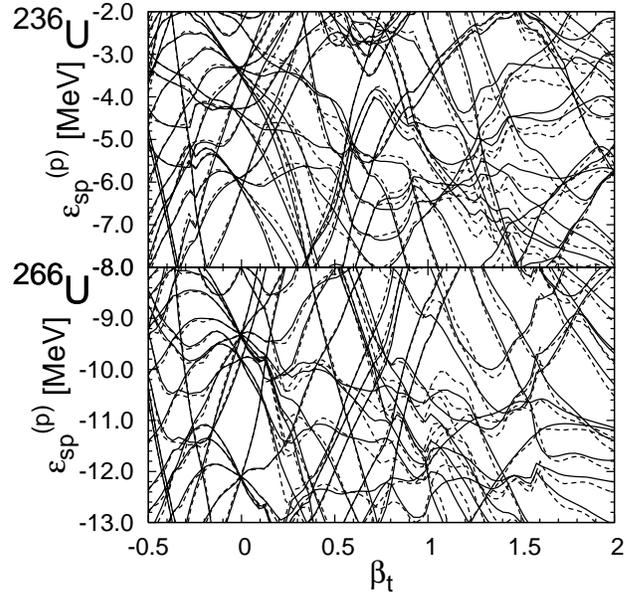}
\caption{
Proton single particle levels near the Fermi energy for  
${}^{236}\rm{U}$ (the top panel) and ${}^{266}\rm{U}$ (the bottom
panel) 
as a function of the total deformation parameter.
The solid and the dashed lines are the results for the total and the 
proton constraints, respectively.
}
\label{fig-spl}
\end{center}
\end{figure}

Figures \ref{fig-fission} and \ref{fig-fission2} show 
the fission barriers for the $^{236}$U and $^{266}$U, respectively,  
as a function of the total deformation parameter $\beta_t$. 
The upper panels are obtained with the total constraint, while the
lower panels show the difference of the fission barrier obtained with
the proton 
constraint from that with the total constraint (the dashed line). 
A similar quantity for the neutron constraint is also shown in the
lower panels by the dotted line. 
The differences are much smaller than the fission barrier height, and 
the fission barriers obtained with the three schemes are almost 
indistinguishable in the scale shown in the figure. 
We have calculated for other even-even uranium isotopes from $^{220}$U
to $^{276}$U, and confirmed that the three schemes lead to almost the
same fission barriers for all of these nuclei. 

Let us next discuss single-particle levels.
Fig. \ref{fig-spl} shows the proton single-particle energies 
near the Fermi energy 
as a function of the total deformation parameter.
The solid and the dashed lines show the results of the total and 
the proton constraints, respectively. 
We see that the single-particle energies are similar to 
each other 
between the total and proton constraints, 
although 
the difference is not negligible at 
large deformations. 
We have found that the tendency is similar also for the neutron 
constraints, although the deviation is smaller as compared to the
proton constraint. 
We have also found that the conclusion remains the same also for the 
neutron single-particle energies. 

\begin{figure}
\begin{center}
\includegraphics[width=0.98\linewidth,clip]{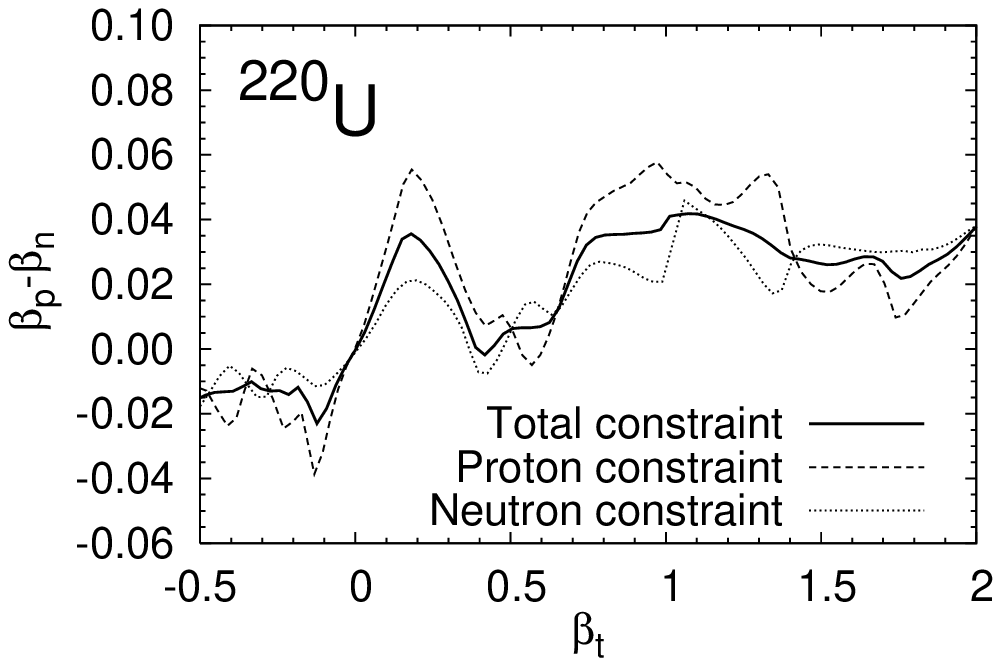}
\includegraphics[width=0.98\linewidth,clip]{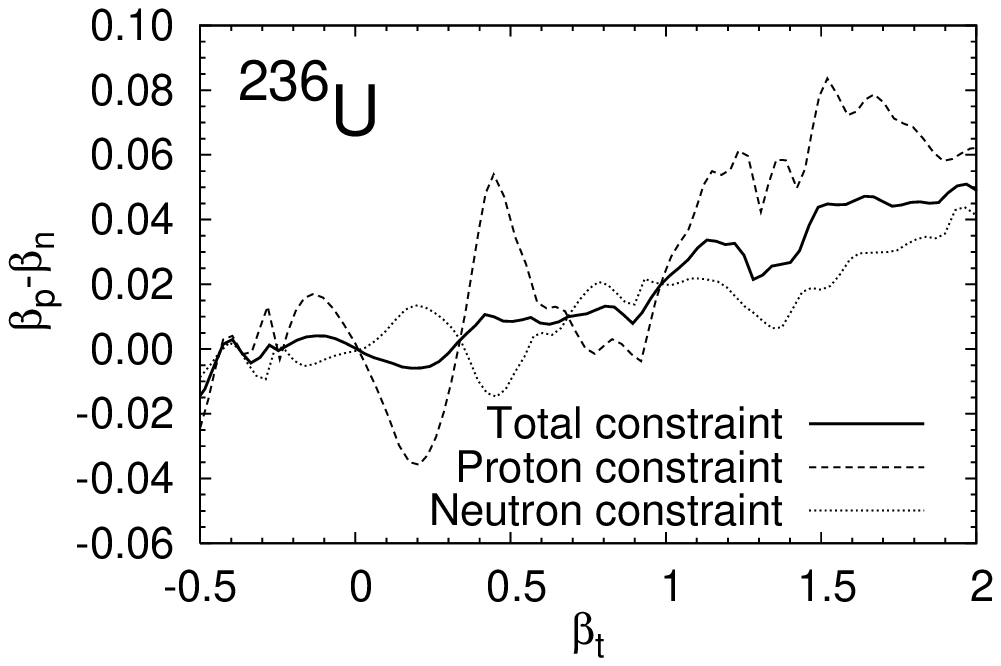}
\includegraphics[width=0.98\linewidth,clip]{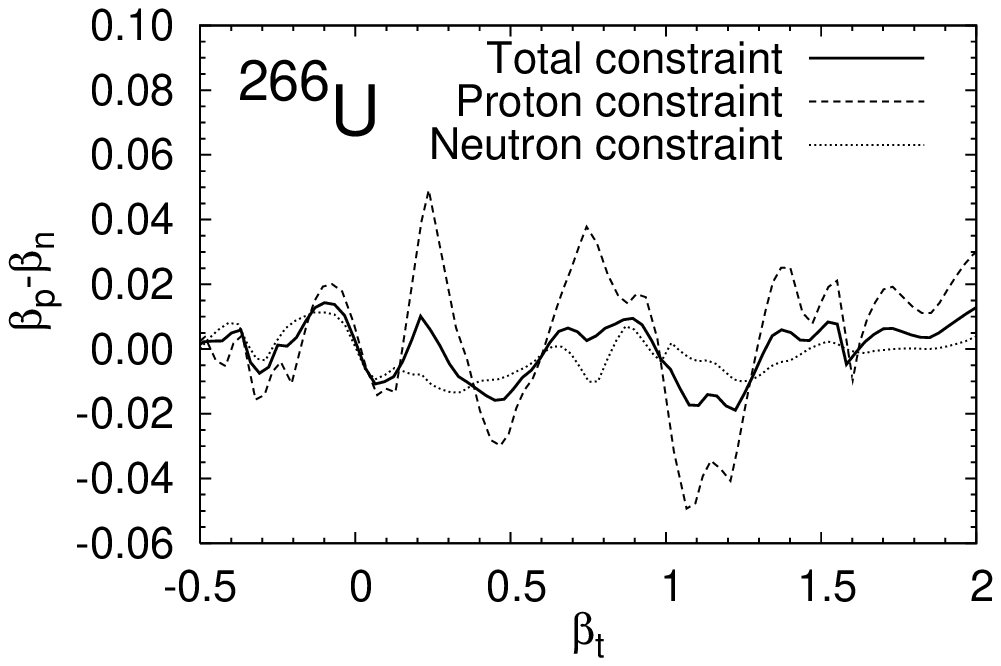}
\caption{
Difference between the proton and the neutron deformation parameters, 
$\beta_p-\beta_n$, for ${}^{220,236,266}\rm{U}$ nuclei  
as a function of the total deformation parameter.
The solid, dashed, and dotted lines are the results 
for the total, proton, and neutron constraints, respectively.
}
\label{fig-beta}
\end{center}
\end{figure}

The difference of deformation parameters for proton and neutron 
along the fission paths 
is shown in Fig. \ref{fig-beta} for ${}^{220,236,266}$U.
Although the difference among the three curves is now more visible 
than in the fission barriers, 
the results with the three schemes are 
similar to each other, indicating that the fission path is not 
sensitive to the constraining operator in the isospin space. 
The $\beta_p-\beta_n$ is not a monotonic function of 
the total deformation parameter $\beta_t$, but on average it 
increases with $\beta_t$ for 
${}^{220}$U and ${}^{236}$U.
Although 
the average value of $\beta_p-\beta_n$ appears to be zero even for large values
of $\beta_t$ for the neutron-rich nucleus ${}^{266}$U, 
this might be an artifact of using the same radius parameter $R_0$ between 
neutron and proton in 
Eq. (\ref{defparameter}) to calculate the deformation parameters (but
see the discussion below). 

\begin{figure}
\begin{center}
\includegraphics[width=0.98\linewidth,clip]{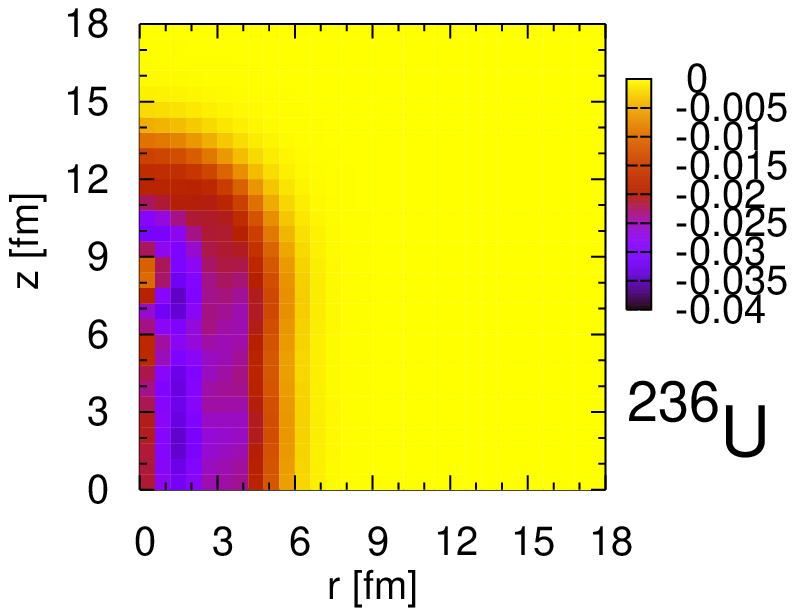}
\includegraphics[width=0.98\linewidth,clip]{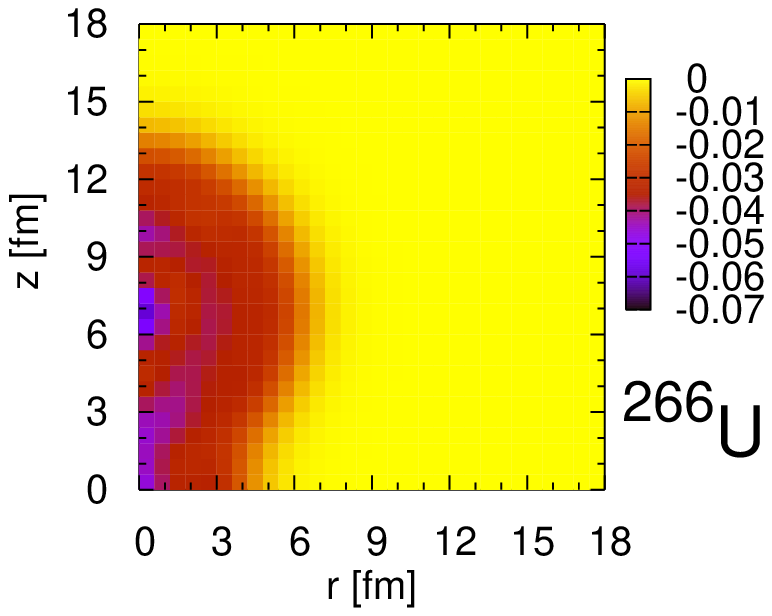}
\caption{(Color online)
Difference of the proton and the 
neutron density distributions, $\rho_p-\rho_n$, 
for ${}^{236}$U (the upper panel) and ${}^{266}$U 
(the lower panel) 
at $\beta_t=2.0$ obtained with the total constraint. 
The densities are axial symmetric around the $z$-axis. 
}
\label{fig-difden}
\end{center}
\end{figure}

Figure \ref{fig-difden} shows the density difference of proton and
neutron, $\rho_p-\rho_n$ for ${}^{236}$U and ${}^{266}$U obtained with 
the total constraint for $\beta_t=2.0$. 
It is plotted 
in the two-dimensional plane of ($r,z$), where the 
density has the axial symmetric shape around the 
$z$ axis. 
One can notice that the difference between the proton and the neutron 
densities is larger in $^{236}$U than in the neutron-rich 
nucleus $^{266}$U. This is consistent with the difference in the 
deformation parameter shown in Fig. \ref{fig-beta}. 

\section{CONCLUSION}

We have used the constrained Skyrme-Hartree-Fock + BCS method 
with a quadrupole constraint 
in order to calculate the fission barriers 
of neutron-rich uranium nuclei with astrophysical interests. 
In particular, we carried out the calculations with the proton,
neutron, and mass (total) quadrupole operators as the 
constraining operators. 
We have found that the fission barriers 
are almost independent of the constraining operators in the
neutron-proton isospin space. 
We have discussed this behaviour using a schematic model, and
suggested 
that the potential energy surface is steep along the isovector degree 
of freedom. 
We have also found that the single-particle levels as well as the 
deformation parameters along the fission paths do not depend much 
on the constraining operators. 

Our calculations indicate that the proton and the neutron deformations 
differ from each other even for the nucleus on the $\beta$-stability 
line, $^{236}$U, and the difference increases as the total deformation 
parameter becomes large. 

In the study of fission barriers based on the mean-field 
approaches, one usually uses mass multipole moments as constraining
operators. Our results suggest that this approach is justified even for 
neutron-rich nuclei, where one might expect that the proton or neutron 
constraint is better. Of course, it is always desirable to treat proton and
neutron separately for neutron-rich nuclei, 
and draw a two-dimensional fission energy surface. 
However, it is rather demanding to do so 
if one has to 
take into account explicitly 
many multipole moments, including simultaneously the quadrupole and octupole
moments, or even higher multipole moments. 
Even in that case, our study clearly indicates that one can reduce the 
number of degree of freedom by introducing the mass multipole
moments, rather than treating proton and neutron separately.

\section*{ACKNOWLEDGMENTS} 
We acknowledge the 21st Century for Center of Excellence (COE) Program 
``Exploring New Science by Bridging Particle-Matter Hierarchy'' at 
Tohoku University for financial support. 
This work was partly supported by the Japanese
Ministry of Education, Culture, Sports, Science and Technology
by Grant-in-Aid for Scientific Research under
the program number 19740115.


\begin{thebibliography}{99}
%
\bibitem{fus06}{\it The proceedings of ``FUSION06: International
  Conference on Reaction Mechanisms and Nuclear Structure at the
  Coulomb barrier''} (edited by L. Corradi, E. Fioretto, A. Gadea, 
D. Ackermann, F. Haas, G. Pollarolo, F. Scarlassara, 
S. Szilner, and M. Trotta), 
AIP Conf. Proc. {\bf 853} (AIP, New York, 2006). 

\bibitem{SFC65}P.A. Seeger, W.A. Fowler, and D.D. Clayton, 
Astrophys. J. Suppl. {\bf 97}, 121 (1965). 
%
\bibitem{Kodama}
T. Kodama, and K. Takahashi,
Nucl. Phys. A \textbf{239}, 489 (1975).
%
%
\bibitem{Role.fis.r-pro.}
G. Martinez-Pinedo, D. Mocelj, N.T. Zinner, A. Keli\'{c}, K. Langanke, I. Panov, B. Pfeiffer, T. Rauscher, K.-H. Schmidt, and F.-K. Thielemann,
Prog. Part. Nucl. Phys. \textbf{59}, 199 (2007).
%
\bibitem{FissionCycling}
J. Beun, G.C. McLaughlin, R. Surman, and W.R. Hix,
Phys. Rev. D \textbf{73}, 093007 (2006); 
arXiv:0707.4498 [astro-ph]. 

\bibitem{Imp.fis.}
I.V. Panov, E. Kolbe, B. Pfeiffer, T. Rauscher, K.-L. Kratz, and F.-K. Thielemann,
Nucl. Phys. A \textbf{747} 633 (2005).

\bibitem{KLF04}E. Kolbe, K. Langanke, and G.M. Fuller, 
Phys. Rev. Lett. {\bf 92}, 111101 (2004). 

\bibitem{moller}P. M\"oller, D.G. Madland, A.J. Sierk, and A. Iwamoto, 
Nature (London) {\bf 409}, 785 (2001); 
P. M\"oller, A.J. Sierk, and A. Iwamoto, Phys. Rev. Lett. {\bf 92},
072501 (2004). 
%
\bibitem{Sys.fis.bar.}
T. B\"{u}rvenich, M. Bender, J. A. Maruhn, and P.-G. Reinhard,
Phys. Rev. C \textbf{69}, 014307 (2004).
%
\bibitem{HFB}
M. Warda, J. L. Egido, L. M. Robledo, and K. Pomorski,
Phys. Rev. C \textbf{66}, 014310 (2002).
%
\bibitem{V.ExtensiontoFission}
M. Samyn, S. Goriely, and J. M. Pearson,
Phys. Rev. C \textbf{72}, 044316 (2005).
%
\bibitem{NWLthesis}
Nyein Wink Lwin, Ph.D.theis,Tohoku University (2007). 
%

\bibitem{QpQnfor16C}
A. P. Severyukhin, M. Bender, H. Flocard, and P.-H. Heenen,
Phys. Rev. C \textbf{75}, 064303 (2007).
%
\bibitem{20O}
T.J. B\"{u}rvenich, Lu Guo, P. Kl\"{u}pfel, P.-G. Reinhard, and W. Greiner,
J. Phys. G: Nucl. Part. Phys. \textbf{35}, 025103 (2008).
%
\bibitem{Dobrowolski}
A. Dobrowolski, K. Pomorski, and J. Bartel
Phys. Rev. C \textbf{75}, 024613 (2007).
%
\bibitem{GHFB}
J. F. Berger and K. Pomorski
Phys. Rev. Lett. \textbf{85}, 30 (2000).
%

\bibitem{VB72}D. Vautherin and D.M. Brink, 
Phys. Rev. C{\bf 5}, 626 (1972). 

\bibitem{BHR03}M. Bender, P.-H. Heenen, and P.-G. Reinhard, 
Rev. Mod. Phys. {\bf 75}, 121 (2003). 

\bibitem{HLY06}K. Hagino, N.W. Lwin, and M. Yamagami, 
Phys. Rev. C{\bf 74}, 017310 (2006). 

\bibitem{skyax}P.-G. Reinhard, the computer code {\tt SKYAX} 
(unpublished). 

\bibitem{RDN99}P.-G. Reinhard, D.J. Dean, W. Nazarewicz,
  J. Dobaczewski, 
J.A. Maruhn, and M.R. Strayer, Phys. Rev. C{\bf 60}, 014316 (1999). 

\bibitem{r-process-path}
S. Goriely and B. Clerbaux,
Astron. Astrophys. \textbf{346}, 798 (1999)

\bibitem{SLy4}
E. Chabanat, P. Bonche, P. Haensel, J. Meyer, R. Schaeffer,
Nucl. Phys. A \textbf{635}, 231 (1998).

\end{thebibliography}
\end{document}